\documentclass[aps,notitlepage,twocolumn,prl,nobalancelastpage,amssymb,superscriptaddress,showpacs,preprintnumbers,nofootinbib,10pt]{revtex4-2}
\usepackage[utf8]{inputenc}
\usepackage[normalem]{ulem}
\usepackage[dvipsnames]{xcolor}
\usepackage{cancel}
\usepackage{amsmath,amsfonts,amsthm,amssymb, bm, mathtools, latexsym} 
\usepackage{subcaption}
\usepackage{physics}
\usepackage[dvipsnames]{xcolor}
\usepackage{enumerate}
\usepackage{slashed}
\usepackage[final]{showlabels}
\usepackage{comment}
\usepackage{aas_macros}
\usepackage{tikz}
\definecolor{red}{rgb}{0.9, 0,0}
\definecolor{cerulean}{rgb}{0., 0.42,0.9}
\definecolor{navy}{rgb}{0.05, 0.05,0.8}

\usepackage[colorlinks]{hyperref}
\hypersetup{
colorlinks = true,
linkcolor = blue,
urlcolor  = blue,
citecolor = purple,
anchorcolor = blue
}

\def\e{{\epsilon}}
\def\ve{{\varepsilon}}
\def\s{{\sigma}}
\def\ka{{\kappa}}
\def\o{{\omega}}
\def\a{{\alpha}}
\def\b{{\beta}}
\def\g{{\gamma}}
\def\l{{\lambda}}
\def\G{{\Gamma}}

\def\d{{\delta}}
\def\t{{\theta}}
\def\D{\Delta}

\def\L{\Lambda}

\def\CA{{\mathcal A}}

\def\CD{{\mathcal D}}

\def\CI{{\mathcal I}}

\def\CK{{\mathcal K}}

\def\CO{{\mathcal O}}

\newcommand{\dt}{{\text d}}
\newcommand{\p}{\partial}

\NewDocumentCommand{\codeword}{v}{%
\texttt{\textcolor{blue}{#1}}%
}

\def\FK{{\text{FK}}}
\def\gr{{\text{gr}}}
\def\ph{{\text{ph}}}

\newcommand{\avg}[1]{\langle #1 \rangle}

\begin{document}

\title{Dressed Fock Spaces in Gauge Theory and Gravity}

\author{Sangmin Choi}
\affiliation{Institute for Theoretical Physics, University of Amsterdam \\
Science Park 904, Postbus 94485, 1090 GL Amsterdam, The Netherlands}
\author{Prahar Mitra}
\affiliation{School of Physics \& Astronomy, University of Southampton, Southampton SO17 1BJ, UK}

\begin{abstract}
Four-dimensional gauge and gravitational theories exhibit long-range interactions that require asymptotic particles to be dressed by clouds of soft photons and gravitons. Faddeev-Kulish dressings render scattering amplitudes infrared-finite, but the resulting multi-particle states do not factorise into tensor products of dressed one-particle states. We show that this loss of Fock-space factorisation is not fundamental, but reflects an inappropriate choice of infrared variables. The real soft divergence is reproduced by the Goldstone modes of asymptotic symmetries, while the Coulomb phase is reproduced by new zero modes of the radiative fields that we introduce here. In these variables, infrared-finite dressed multi-particle states admit the usual Fock-space factorisation into single-particle dressed states.
\end{abstract}

\maketitle

\section{Introduction}
\label{sec:intro}

In the scattering problem, one considers a quantum system whose Hamiltonian $H$ can be decomposed into a free part $H_0$ and an interaction $V$, $H = H_0 + V$. The scattering amplitude is then defined as
\begin{equation}
\begin{split}
\label{Sfi_def}
S_{\b\a}  = \lim_{T \to \infty} \bra{\b} e^{ i H_0 T } e^{- i H (2T) } e^{  i H_0 T } \ket{\a} .
\end{split}
\end{equation}
In the standard treatment, the initial and final states are taken to be collections of particles that are widely separated at early and late times. These are well approximated by non-interacting multi-particle eigenstates of $H_0$, namely Fock states,
\begin{equation}
\begin{split}
\label{gapped_assumption}
\bra{\b} &= \bra{0} \CO^+_1(p_1) \cdots \CO^+_m(p_m)  , \\
\ket{\a} &= \CO^-_{m+1}(p_{m+1}) \cdots \CO^-_n ( p_n ) \ket{0}  , \\
\end{split}
\end{equation}
where $\CO_i^+(p_i)$ ($\CO_i^-(p_i)$) is an annihilation (creation) operator for the $i$th particle. We suppress spin, helicity, internal quantum numbers, Bose/Fermi symmetrisation, and the use of normalisable wavepackets, none of which will play an essential role in this paper. We will refer to the scattering amplitude for such states as the Fock amplitude.

The description above is appropriate in theories with a mass gap. It fails, however, in four-dimensional gauge and gravitational theories, where photons and gravitons mediate long-range interactions that persist even when particles are far-separated. In such theories, the Fock amplitude vanishes due to infrared divergences \cite{mott_1931, Bloch:1937pw, Yennie:1961ad, Weinberg:1965aa}. This problem was investigated in QED by Faddeev and Kulish (FK) \cite{Kulish:1970ut} (building on earlier work of Chung \cite{Chung:1965zza}, Kibble \cite{Kibble:1968sfb, Kibble:1968aa, Kibble:1968ab, Kibble:1968ac}, and Dollard \cite{Dollard:1964cok}), and in gravity by Ware, Saotome and Akhoury \cite{Ware:2013zja}. These works showed that infrared-finite amplitudes can be obtained by replacing ordinary Fock states with dressed asymptotic states of the form
\begin{equation}
\begin{split}
\label{FK_states}
\,_\FK\bra{\b} &= \bra{\b} Z^+_\b , \qquad \ket{\a}_\FK = Z^-_\a \ket{\a}. 
\end{split}
\end{equation}
Scattering amplitudes evaluated in the FK states are finite and non-vanishing \cite{Chung:1965zza, Kulish:1970ut, Choi:2017bna}. The dressing factor $Z$ is a unitary operator that has the form $e^R e^{i\Phi}$. $R$ is linear in the soft (low-energy) photon and graviton creation and annihilation operators and linear in the charge- or energy-density operator, so the operator $e^R$ produces a coherent cloud of soft photons and gravitons. In particular, this implies that FK states have finite energy, but do not have a well-defined particle number. On the other hand, the operator $\Phi$ is quadratic in the charge- or energy-density operator, so that the ``Coulomb phase operator'' $e^{i\Phi}$ acts on pairs of one-particle states.\footnote{In the language of quantum optics, $e^R$ is a \emph{displacement operator} for soft photons and gravitons and a \emph{phase-shift operator} for matter particles. $e^{i\Phi}$ is a \emph{multi-mode cross-Kerr operator} for matter particles.} This implies that a multi-particle FK state does not factorise into a tensor product of one-particle dressed states, e.g.
\begin{equation}
\begin{split}
\label{FK_non_factor}
\,_\FK\!\bra{\b} &\neq \bra{0} {\tilde \CO}^+_1(p_1) \cdots {\tilde \CO}^+_m(p_m) , \\
\ket{\a}_\FK &\neq {\tilde \CO}^-_{m+1}(p_{m+1}) \cdots {\tilde \CO}^-_n ( p_n ) \ket{0} .
\end{split}
\end{equation}
This lack of factorisation of FK states obscures several basic properties of dressed amplitudes. For example, it is not known in general how to formulate an LSZ reduction formula directly for dressed amplitudes (see \cite{Dybalski:2017mip} for related work); and the factorisation of dressed amplitudes on physical poles is subtle (see \cite{Lippstreu:2025jit} for related work).

These observations motivate the central question of this paper: is the failure of Fock-space factorisation a fundamental feature of four-dimensional gauge and gravitational theories, or is it a consequence of the particular framework used to describe the infrared sector? In this paper, we reformulate the dressing so that the infrared-divergent factor of the amplitude is captured by single-particle operators acting on an enlarged asymptotic Hilbert space. This space contains both hard particle degrees of freedom and zero-energy soft modes, and the hard and soft sectors commute. In this description, ordinary asymptotic creation and annihilation operators admit a soft-hard factorisation
\begin{equation}
\begin{split}
\label{Oi_factorisation}
\CO^\pm_i(p_i) = U_i^\pm(p_i) {\tilde \CO}^\pm_i(p_i) , 
\end{split}
\end{equation}
where $U^\pm_i$ acts only on the soft sector, while ${\tilde \CO}^\pm_i$ creates or annihilates the corresponding hard particle in the dressed asymptotic Hilbert space. We show that correlators of $U^\pm_i$ reproduce the infrared-divergent factors of the ordinary Fock amplitude. It follows that the scattering amplitude evaluated in the tensor product multi-particle states
\begin{equation}
\begin{split}
\label{dressed_def}
\,_{\text{D}}\!\bra{\b} &= \bra{0} {\tilde \CO}^+_1(p_1) \cdots {\tilde \CO}^+_m(p_m) , \\
\ket{\a}_{\text{D}} &= {\tilde \CO}^-_{m+1}(p_{m+1}) \cdots {\tilde \CO}^-_n ( p_n ) \ket{0}
\end{split}
\end{equation}
is infrared finite. These states can be expressed in terms of the Fock states \eqref{gapped_assumption} as
\begin{equation}
\begin{split}
\label{D_states}
\,_\text{D}\!\bra{\b} &= \bra{\b} {\tilde Z}^+_\b , \qquad \ket{\a}_\text{D} = {\tilde Z}^-_\a \ket{\a} , 
\end{split}
\end{equation}
where
\begin{equation}
\begin{split}
{\tilde Z}_\b^+ = ( U_1^+ \cdots U_m^+ )^{-1} , \qquad {\tilde Z}_\a^- = ( U_{m+1}^- \cdots U_n^- )^{-1} .
\end{split}
\end{equation}
In essence, the many-body FK dressing \eqref{FK_states} is replaced here by a product of single-particle soft operators whose contractions reproduce the same infrared phase.

The asymptotic Hilbert space comprising states of the form \eqref{dressed_def} therefore admits a Fock-like tensor-product representation, and the associated amplitude is infrared finite and non-vanishing. In particular, the construction preserves the advantages of FK dressing while restoring the tensor product structure needed for standard notions of asymptotic particles, factorisation, and scattering amplitudes. This suggests that infrared-finite scattering in four-dimensional gauge theories and gravity can be formulated in a way that is fully compatible with a Fock-space description, provided the soft sector is represented appropriately.

To construct the operators $U_i$, we draw on recent advances in our understanding of the asymptotic structure of gauge and gravitational theories in asymptotically flat spacetimes \cite{Strominger:2017zoo, Pasterski:2021rjz, Raclariu:2021zjz}. These theories admit an infinite-dimensional asymptotic symmetry group --- large gauge transformations for gauge theories and supertranslations for gravity \cite{He:2014cra, He:2014laa} --- that is spontaneously broken down to the global $U(1)$ in gauge theories and the four translations in gravity. The corresponding Goldstone modes $C_\ph$ and $C_\gr$ are scalar fields that live on the celestial sphere. They provide a natural boundary description of the soft sector. In \cite{Kapec:2017tkm, Arkani-Hamed:2020gyp, Kalyanapuram:2021bvf, Kapec:2021eug, He:2024ddb, He:2024skc, Choi:2026dhz}, it was shown that part of the operator $U_i$ -- the one that reproduces the real part of the infrared divergence -- can be expressed in terms of these Goldstone modes. It does not, however, account for the infrared-divergent phase, often called the Coulomb phase. In the traditional FK construction, this phase is reproduced by the dressing $e^{i\Phi}$, and it is precisely this operator that obstructs the tensor product structure of the asymptotic Hilbert space. The main result of this paper is that the Coulomb phase can \emph{also} be represented in terms of soft degrees of freedom, but \emph{not} in terms of the aforementioned Goldstone modes. Instead, we uncover two \emph{new} zero-energy scalar modes $N^\pm$ that live on the three-dimensional hyperboloid $H_3$ that arises in the de Boer-Solodukhin description of the blow-up of $i^\pm$ \cite{deBoer:2003vf, Campiglia:2015qka}, and show that the Coulomb phase can be reproduced by dressings built out of these zero modes.

The paper is organised as follows. In section \ref{sec:irdiv}, we review infrared divergences in four-dimensional Abelian gauge theories, and also review how the real infrared divergence is reproduced by a Goldstone dressing. In section \ref{sec:Npm}, we introduce the zero modes $N^\pm_\ph$ and derive their two-point functions. In section \ref{sec:Nphdef}, we determine the contribution of these zero modes to the infrared dressing and show that it reproduces the Coulomb phase. We use these modes to construct infrared-finite dressed states that factorise into tensor products of one-particle states. In section \ref{sec:gravity}, we extend the construction to perturbative gravity. In section \ref{sec:massless}, we discuss the massless limit. Finally, we conclude in section \ref{sec:conclusions} with comments on the implications and future directions.

\section{Infrared Divergences}
\label{sec:irdiv}

Consider an $n$-point Fock amplitude $\CA_n$ consisting of $n$ particles with on-shell momenta $p_i^\mu$ ($p_i^2 = - m_i^2$ and $p_i^0>0$) and electric charge $Q_i \in {\mathbb Z}$. Infrared divergences in the amplitude are regulated by a cutoff $\mu$. The integration regions of loop photons are restricted to $|\vec{\ell}\,|  > \mu$. In the limit $\mu \ll \Lambda \ll E$ where $E$ is the typical energy scale of the external particles, the amplitude at scale $\mu$ is related to the amplitude at scale $\Lambda$ by (see \cite{Weinberg:1995mt})
\begin{equation}
\begin{split}
\label{soft_factorisation_ID}
\CA_n(\mu) \equiv \avg{ \CO^{\eta_1}_1(p_1) \cdots  \CO^{\eta_n}_n(p_n)} = e^{-\G_\ph } \CA_n(\L)
\end{split}
\end{equation}
where\footnote{When $i=j$, we replace $ - p_j \cdot \ell - i \e \to  - p_i \cdot \ell + i \e$.}
\begin{equation}
\begin{split}
\label{Gamma_ph_def}
\G_\ph &= \frac{i e^2}{2} \sum_{ij} \eta_i \eta_j Q_i Q_j \int_\mu^\L \frac{\dt^4\ell}{(2\pi)^4} \\
&\qquad \times \frac{p_i \cdot p_j }{  ( \ell^2 - i\e ) (  p_i \cdot \ell - i \eta_i \e ) (  - p_j \cdot \ell - i \eta_j \e )}.
\end{split}
\end{equation}
Here $e$ is the dimensionless coupling constant of the gauge theory, $\eta_i=+1$ for outgoing particles and $\eta_i=-1$ for incoming particles. The integral has been explicitly evaluated in \cite{Weinberg:1995mt} and we simply present the result here. For now, we assume that all charged particles are massive. Massless particles are discussed later in section \ref{sec:massless}. For massive particles, we define a rescaled vector
\begin{equation}
\begin{split}
\label{Pi_def}
P_i^\mu = \frac{p_i^\mu}{m_i} \quad \implies \quad  P_i^2 = - 1 , \qquad P_i^0 > 0 . 
\end{split}
\end{equation}
We see that $P_i$ is a point on a three-dimensional Euclidean hyperboloid $H_3$. The geodesic distance between two points on $H_3$ is $\s_{ij} = \cosh^{-1} ( - P_i \cdot P_j )$. In terms of this,\footnote{In \cite{Weinberg:1995mt}, the result is expressed in terms of $\b_{ij} = \tanh \s_{ij}$.} we find
\begin{equation}
\begin{split}
\label{Gamma_ph_calc}
\G_\ph &= - \frac{e^2}{8\pi^2} \ln \frac{\L}{\mu} \sum_{ij}  \eta_i \eta_j Q_i Q_j \s_{ij} \coth \s_{ij} \\
&\qquad \qquad + \frac{i e^2}{8\pi}  \ln \frac{\L}{\mu}  \sum_{i \neq j} \d_{\eta_i,\eta_j} Q_i Q_j\coth \s_{ij} .
\end{split}
\end{equation}
Note that as $\mu \to 0$, we have $\Re \G_\ph \to + \infty$ which implies that the Fock amplitude vanishes when the regulator is removed, $\CA_n(0)=0$. This is the aforementioned infrared divergence of Fock amplitudes.

The goal of this paper is to construct the operators $U_i^\pm$ such that
\begin{equation}
\begin{split}
\avg{ U_1^{\eta_1} \cdots U_n^{\eta_n} } = e^{ - \G_\ph } .
\end{split}
\end{equation}
It then follows from \eqref{Oi_factorisation} and \eqref{soft_factorisation_ID} that
\begin{equation}
\begin{split}
\avg{ {\tilde \CO}^{\eta_1}_1(p_1) \cdots {\tilde \CO}^{\eta_n}_n(p_n)} = \CA_n(\L) .
\end{split}
\end{equation}
Note that the infrared cutoff $\mu$ has completely dropped out of the dressed amplitude, so it is infrared finite.\footnote{The dressed amplitude still depends on the cutoff $\L$. This can be interpreted as the soft-hard factorisation scale, or equivalently as the energy resolution of the detector. The same energy scale also enters in the soft operators $U_i^{\eta_i}$.}

Following the work of Faddeev and Kulish \cite{Kulish:1970ut}, it is convenient to decompose this operator as
\begin{equation}
\begin{split}
\label{U_decomp}
U_i^\pm = e^{{\tilde R}^\pm_i} {:} e^{i {\tilde \Phi}^\pm_i}{:} .
\end{split}
\end{equation}
The operators ${\tilde R}^\pm_i$ and ${\tilde \Phi}^\pm_i$ are decoupled and satisfy
\begin{equation}
\begin{split}
\label{U_decomp_prop}
\avg{ \prod_i e^{{\tilde R}^{\eta_i}_i} } = e^{ - \Re \G_\ph } , \quad \avg{  \prod_i {:} e^{i {\tilde \Phi}^{\eta_i}_i} {:} } = e^{ - i \Im \G_\ph } . 
\end{split}
\end{equation}
The operator ${\tilde R}^\pm_i$ has been constructed previously \cite{Himwich:2020rro, Kapec:2021eug, Magnea:2021fvy, Gonzalez:2021dxw, Kalyanapuram:2021bvf, He:2024ddb, He:2024skc} and we review these works now. ${\tilde \Phi}^\pm_i$ and the normal-ordering are discussed in the next section.

Abelian gauge theories are invariant under $U(1)$ gauge transformations,
\begin{equation}
\begin{split}
\label{gauge_symmetry}
A_\mu(x) \to A_\mu(x) + \p_\mu \l(x) , \qquad \l(x) \sim \l(x) + 2\pi . 
\end{split}
\end{equation}
To describe this symmetry, it is convenient to decompose the gauge field as
\begin{equation}
\begin{split}
\label{A_decomp}
A_\mu(x) = {\hat A}_\mu(x) + \p_\mu C(x)  
\end{split}
\end{equation}
where ${\hat A}_\mu(x)$ is defined in terms of the field strength, $F_{\mu\nu}(x) = \p_\mu A_\nu(x) - \p_\nu A_\mu(x) $. The precise form is fixed by a choice of gauge. In this paper, we work in Fock-Schwinger gauge,
\begin{equation}
\begin{split}
\label{FS_gauge}
x^\mu {\hat A}_\mu(x) = 0 \implies {\hat A}_\mu(x) = - x^\nu \int_0^1 \dt s\, s F_{\mu\nu}(s x) . 
\end{split}
\end{equation}
The zero mode $C(x)$ appears in \eqref{A_decomp} through a derivative so it is only defined up to a constant additive shift
\begin{equation}
\begin{split}
\label{C_identification}
C(x) \sim C(x) + c . 
\end{split}
\end{equation}
The gauge transformation \eqref{gauge_symmetry} then acts as
\begin{equation}
\begin{split}
\label{gauge_symmetry_1}
{\hat A}_\mu(x) \to {\hat A}_\mu(x) , \qquad C(x) \to C(x) + \l(x) .
\end{split}
\end{equation}
Notice that due to the identification \eqref{C_identification}, $C(x)$ transforms non-trivially only if $\l(x)$ is a non-constant function. It does \emph{not} transform under global $U(1)$ transformations.

For $\l(x)$ that vanishes at spatial infinity, \eqref{gauge_symmetry_1} represents a redundancy in our description of the system. On the other hand, for $\l(x)$ that limits to a finite function $\ve({\hat q})$ at infinity, this is a physical symmetry known as \emph{large gauge transformation} \cite{He:2014cra}. It was shown in \cite{He:2014cra, Campiglia:2015qka} that the Ward identity associated with this symmetry is Weinberg's leading soft-photon theorem \cite{Weinberg:1965aa}. Here, we use the standard embedding space notation to denote a point on the celestial sphere (${\hat q}^2 = 0$ and ${\hat q}^\mu \sim \l {\hat q}^\mu$). In the Fock basis, the large gauge symmetry is spontaneously broken down to the global $U(1)$ symmetry $\ve({\hat q}) \to \ve_0$. The associated Goldstone mode is denoted by $C_\ph({\hat q})$, which is the value of $C(x)$ at spatial infinity. It was shown in \cite{Himwich:2020rro, Kalyanapuram:2021bvf, Kapec:2021eug} that $C_\ph({\hat q})$ is a Gaussian operator with
\begin{equation}
\begin{split}
\label{C_2pt}
\avg{ C_\ph({\hat q}) C_\ph({\hat q}') } = - \frac{e^2}{4\pi^2} \ln \frac{\L}{\mu} \ln ( - {\hat q} \cdot {\hat q}' ) .
\end{split}
\end{equation}
This is the two-point function of a free boson in ${\mathbb R}^2$.

The dressing $e^{{\tilde R}_i^\pm}$ is constructed using this Goldstone mode as follows. It was shown in \cite{Mandelstam:1962mi,Jakob:1990zi,Feige:2013zla,Feige:2014wja} that charged particles in a scattering amplitude are dressed with a Wilson line
\begin{equation}
\begin{split}
\label{Wilson_Line}
W_i = \exp \left( i Q_i \int_{\g_i} A \right) ,
\end{split}
\end{equation}
where $\g_i$ is the worldline of the $i$th particle. The dressing responsible for the real part of the infrared divergence is obtained by extracting the contribution of $C_\ph({\hat q})$ to this Wilson line. To do this, we substitute \eqref{A_decomp} into \eqref{Wilson_Line} and extract the contribution of $C(x)$ (the contribution of ${\hat A}$ will be discussed in section \ref{sec:Nphdef}). This is\footnote{All the worldlines $\g_i$ start at the origin $x^\mu=0$ and extend out to infinity. Consequently, in addition to the boundary at infinity, there is also a boundary at the origin, which contributes $-i\eta_i Q_i C(0)$. This contribution vanishes in the product $\prod_i e^{ {\tilde R}^{\eta_i}_i}$ due to total charge conservation.}
\begin{equation}
\begin{split}
\label{Ri_def}
{\tilde R}_i^{\eta_i} &=  i Q_i \int_{\g_i} \dt C = i Q_i C |_{\p \g_i} = i \eta_i Q_i C^{\eta_i}(P_i) ,
\end{split}
\end{equation}
where $C^\pm = C|_{i^\pm}$. The sign $\eta_i=\pm1$ arises from the fact that $i^+$ is a future boundary whereas $i^-$ is a past boundary. Here, we use the fact that the endpoint of the worldline $\g_i$ on the hyperbolic blow-up of $i^\pm$ is $P_i^\mu$ \cite{Campiglia:2015qka}. It was further shown in that work that $C^{\eta_i}(P_i)$ can be written in terms of the Goldstone mode $C_\ph({\hat q})$ as\footnote{There is no $\pm$ superscript on the Goldstone mode $C_\ph({\hat q})$ due to the antipodal matching conditions \cite{He:2014cra, Campiglia:2017mua, Capone:2022gme}.}
\begin{equation}
\begin{split}
\label{CPi_def}
C^{\eta_i}(P_i) \equiv \int  \dt^2 {\hat q}\, \CK_2 ( P_i , {\hat q} ) C_\ph ({\hat q})  
\end{split}
\end{equation}
where $\CK_\D(P,{\hat q})$ is the scalar bulk-boundary propagator in $H_3$,
\begin{equation}
\begin{split}
\label{Bb_def}
\CK_\D ( P , {\hat q} ) = \frac{\D-1}{\pi ( - 2 P \cdot {\hat q} )^\D } . 
\end{split}
\end{equation}
Note that due to the identification \eqref{C_identification}, the dressing is only defined up to ${\tilde R}^{\eta_i}_i \sim {\tilde R}^{\eta_i}_i +  i c \eta_i Q_i$. However, the total dressing over all particles $\sum_i {\tilde R}^{\eta_i}_i$ is well-defined due to total charge conservation.

With this, we have all the ingredients to perform our calculation. Using the dressing \eqref{Ri_def} with \eqref{CPi_def}, and using the following property that applies to Gaussian operators with zero mean
\begin{equation}
\begin{split}
\label{Gaussian_Property}
\avg{ e^{ i \int \dt x \phi(x) J(x) } } &= e^{ - \frac{1}{2} \int \dt x \dt x'  \avg{ \phi(x) \phi(x') }  J(x) J(x') } ,
\end{split}
\end{equation}
a straightforward computation gives
\begin{equation}
\begin{split}
\avg{ \prod_i e^{{\tilde R}^{\eta_i}_i} } = e^{ - \Re \G_\ph } 
\end{split}
\end{equation}
which is the first equation in \eqref{U_decomp_prop}.

\section{A New Zero Mode}
\label{sec:Npm}

In this section, we introduce two new zero modes $N^\pm$ and determine their two-point function. We start by considering the two-point function of these modes in the free theory, which is described by the bulk Maxwell action,
\begin{equation}
\begin{split}
\label{Maxwell_Action}
S  = - \frac{1}{4e^2} \int_{M_4} \dt^4 x\, F^{\mu\nu} F_{\mu\nu}  + S_{\rm bdy} . 
\end{split}
\end{equation}
We now substitute the decomposition \eqref{A_decomp} into this action. The pure gauge mode $C(x)$ does not contribute to the bulk Maxwell action \eqref{Maxwell_Action}, but it does contribute to the boundary action $S_{\rm bdy}$. This was used in \cite{He:2024ddb} to construct the so-called \emph{soft effective action} \cite{Kalyanapuram:2021bvf, Nguyen:2021ydb, Kapec:2021eug} from which the two-point function \eqref{C_2pt} follows. In this section, we are interested in the zero modes $N^\pm_\ph$ that are encoded in the radiative field ${\hat A}_\mu(x)$.

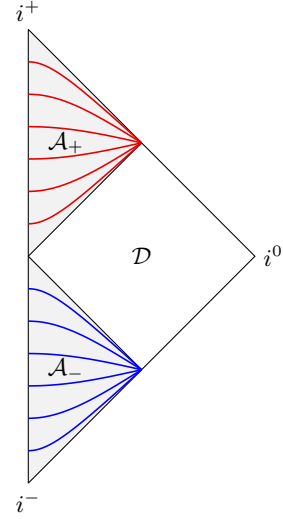
\begin{figure}[htbp!]
\centering
\begin{tikzpicture}[scale=3]
\coordinate (O) at (0,0);
\coordinate (S) at (0,-1);
\coordinate (N) at (0,1);
\coordinate (E) at (1,0);
\coordinate (W) at (-1,0);
\coordinate (SW) at (-1/2,-1/2);
\coordinate (SE) at (1/2,-1/2);
\coordinate (NW) at (-1/2,1/2);
\coordinate (NE) at (1/2,1/2);
\fill[gray!10] (O) -- (N) -- (NE) -- cycle;
\fill[gray!10] (O) -- (SE) -- (S) -- cycle;
\draw (N) -- (E) -- (S);
\draw (N) -- (S);
\draw (O) -- (SE);
\draw (NE) -- (O);
\node[right] at (E) {$i^0$};
\node[above] at (N) {$i^+$};
\node[below] at (S) {$i^-$};
\node[font=\small] at (1/6,1/2) {$\CA_+$};
\node[font=\small] at (1/6,-1/2) {$\CA_-$};
\node[font=\small] at (1/2,0) {$\CD$};
\def\Nt{6}
\def\Nr{10}
\tikzset{declare function={
T(\t,\r) = {\fpeval{(atan(\t*\r+\t*sqrt(1+\r^2))-atan(\t*\r-\t*sqrt(1+\r^2)))/pi}};
R(\t,\r) = {\fpeval{(atan(\t*\r+\t*sqrt(1+\r^2))+atan(\t*\r-\t*sqrt(1+\r^2)))/pi}};
}}
\foreach \i [evaluate={\t=tan(90*\i/(\Nt+1))}] in {1,...,\Nt}{
\draw[
color=red,
line width=.6,
samples=\Nr,
smooth,
variable=\r,
domain=0.001:0.999
]
plot ({R(\t,tan(90*\r))},{T(\t,tan(90*\r))});
\draw[
color=blue,
line width=.6,
samples=\Nr,
smooth,
variable=\r,
domain=0.001:0.999
]
plot ({R(\t,tan(90*\r))},{-T(\t,tan(90*\r))});
}
\end{tikzpicture}
\caption{Penrose diagram of Minkowski spacetime. The upper and lower shaded regions, denoted by $\CA_+$ and $\CA_-$, correspond respectively to the interiors of the future and past light cones. Coloured curves represent constant $\tau$ hypersurfaces. The complementary region is denoted by $\CD$.
}
\label{fig:Penrose}
\end{figure}

To describe $i^\pm$, we use the construction of de Boer and Solodukhin in \cite{deBoer:2003vf}. We decompose Minkowski space $M_4$ into three regions demarcated by the light cone centred at $x^\mu=0$, see Figure~\ref{fig:Penrose}. In the interiors of the future ($+$) and past ($-$) light cones $\CA_\pm$, we set up coordinates
\begin{equation}
\begin{split}
\label{hyperbolic_coord}
x^\mu = \pm e^\tau P^\mu , \qquad P^2 = - 1 , \quad P^0 > 0 , \qquad \tau \in {\mathbb R}.
\end{split}
\end{equation}
$i^\pm$ is reached by taking $\tau \to \infty$, keeping $P$ fixed. The opposite limit ($\tau \to - \infty$) zooms into the origin $x^\mu=0$. These coordinates cover the region $x^2 < 0$ of Minkowski spacetime, and describe the blow-ups of $i^\pm$. The complementary region $\CD: x^2 > 0$ is used to describe the blow-up of spatial infinity $i^0$ and will not play a role in this paper.

\eqref{FS_gauge} implies ${\hat A} |_{\CA_\pm} = {\hat A}^\pm_a \dt y^a $, where $y^a$ are the intrinsic coordinates on $H_3$. Note also that ghosts decouple from the theory in this gauge. From \eqref{FS_gauge}, we see that ${\hat A}$ vanishes at the origin,
\begin{equation}
\begin{split}
\label{regularity_cond}
{\hat A}_a^\pm(\tau)  = \CO(e^{2\tau}) \quad \text{as $\tau \to -\infty$.}
\end{split}
\end{equation}
We further restrict our attention to configurations that are finite near $i^\pm$,\footnote{In the presence of charged matter, ${\hat A}_a^\pm$ diverges linearly in $\tau$ as $\tau \to \infty$. The coefficient of this linear divergence is entirely determined by the charged matter fields and is not a dynamical mode \cite{Campiglia:2019wxe}.}
\begin{equation}
\begin{split}
\label{asymp_cond}
{\hat A}_a^\pm(\tau)  = N_a^\pm + \CO(e^{-2\tau}) \quad \text{as $\tau \to +\infty$.}
\end{split}
\end{equation}
$N_a^\pm$ is the zero mode that we are interested in. This can be described in terms of the field strength as
\begin{equation}
\begin{split}
\label{Na_field_rel}
N_a^\pm = \int_{\mathbb R} \dt \tau \,F^\pm_{\tau a} .
\end{split}
\end{equation}
It follows from this that $N_a^\pm$ has zero Milne energy \cite{Cheung:2016iub}.

To derive an effective action for $N_a^\pm$, it is convenient to move to a Fourier basis
\begin{equation}
\begin{split}
\label{Ahat_Fourier}
{\hat A}_a^\pm (\tau)  =  \int_{\mathbb R} \frac{\dt \nu}{2\pi i} \frac{ e^{ i  \nu \tau } }{\nu - i 0^+} {\tilde A}^\pm_a (\nu) .
\end{split}
\end{equation}
The boundary conditions \eqref{regularity_cond} and \eqref{asymp_cond} imply that the Fourier transform of ${\hat A}_a^\pm(\tau)$ has a pole at $\nu = i0^+$. This has been accounted for in \eqref{Ahat_Fourier} so ${\tilde A}^\pm_a (\nu)$ is finite as $\nu \to 0$ with ${\tilde A}^\pm_a (0) = N_a^\pm$. In terms of the Fourier coefficient, the action is
\begin{equation}
\begin{split}
\label{action_Anu}
 S^\pm  &= \frac{1}{e^2}  \int_{\mathbb R}  \frac{\dt \nu}{2\pi} \frac{1}{\nu^2} \! \int_{H_3} \!\!\! \dt^3 P \! \left(  \! - \frac{1}{4} |{\tilde F}^\pm_{ab}|^2 + \frac{1}{2} \nu^2 |{\tilde A}^\pm_a|^2 \! \right)  \! .
\end{split}
\end{equation}
This is the action for infinitely many (complex) vector fields on $H_3$ labelled by $\nu$ with mass $m^2_\nu = - \nu^2$. Using the AdS/CFT dictionary $(\D-1)^2 = m^2$, we find that the boundary operator has dimension $\D = 1 \pm i \nu$ (the sign choice depends on the boundary conditions at $\p H_3$).  From the action, we can extract the two-point function \cite{Allen:1985wd}
\begin{equation}
\begin{split}
\label{2pt_gen}
&\avg{ {\tilde A}^\pm_a(\nu,P) {\tilde A}^\pm_b(\nu',P') } = \frac{ie^2}{2}  \d(\nu+\nu') \\
&\qquad \qquad \times [ \p_a \p'_b + \nu^2 \p_a \s \p'_b \s ]  [ e^{ i \nu \s } ( \coth \s - i \nu ) ]  .
\end{split}
\end{equation}
This is derived by solving the Schwinger-Dyson equation that follows from the action \eqref{action_Anu}. This is a second-order differential equation. One of the integration constants is fixed by requiring that the coincident point singularity $\s \to 0$ matches the one in flat spacetime. The other integration constant is fixed by Neumann boundary conditions on $\p H_3$.\footnote{For every $\tau$, $\p H_3$ is the celestial sphere on $\CI^\pm$ (see Figure~\ref{fig:Penrose}). The Neumann boundary condition corresponds to allowing radiative flux through the celestial boundary.}

Setting $\nu=\nu'=0$, we find that the two-point function of the zero mode $N_a^\pm$ is given by
\begin{equation}
\begin{split}
\label{Na_2pt}
\avg{ N^\pm_a(P) N^\pm_b(P') } &= \frac{ie^2}{2}  \d(0)  \p_a \p'_b \coth \s . 
\end{split}
\end{equation}
From this, it follows that $N^\pm_a$ is purely longitudinal and we can write
\begin{equation}
\begin{split}
\label{Na_der}
N^\pm_a \equiv \p_a N^\pm_\ph .
\end{split}
\end{equation}
Note that, just like the zero mode $C(x)$ \eqref{C_identification}, the scalar mode $N^\pm_\ph$ is only defined up to additive shifts,
\begin{equation}
\begin{split}
\label{N_identification}
N^\pm_\ph(P) \sim N^\pm_\ph(P) + c.
\end{split}
\end{equation}
The two-point function of $N^\pm_\ph$ can be easily extracted from \eqref{Na_2pt}. Before we do that, let us discuss the divergent factor $\d(0)$. This factor arises from the infinite volume of $\CA_\pm$, as $\d(0) = \frac{1}{2\pi} \int_{\mathbb R} \dt \tau$ and is therefore related to the infrared divergences in the gauge theory. In the previous section, we regulated this infrared divergence by introducing the cutoffs $\mu$ and $\Lambda$. These same cutoffs must therefore also regulate $\d(0)$. Since the cutoffs are imposed in energy space, we need to relate the parameter $\nu$ to $\o$. To do this, we exploit the now standard relationship between the scaling dimension $\D=1+i\nu$ and energy $\o$ given by the Mellin transform
\begin{equation}
\begin{split}
2\pi \d(\nu-\nu') =  \int_0^\infty \dt \o\, \o^{i(\nu-\nu')-1} .
\end{split}
\end{equation}
Setting $\nu=\nu'$, we find
\begin{equation}
\begin{split}
\d(0) =  \frac{1}{2\pi} \int_0^\infty \frac{\dt \o}{\o} .
\end{split}
\end{equation}
The divergent integral is precisely the logarithmic infrared divergence in gauge theories. We use the cutoffs to regulate this $\int_0^\infty \to \int_\mu^\L$, and find 
\begin{equation}
\begin{split}
\d_\text{reg}(0) =  \frac{1}{2\pi} \ln \frac{\L}{\mu}  .
\end{split}
\end{equation}
With all this, we find that the longitudinal mode is Gaussian, and its connected two-point function is
\begin{equation}
\begin{split}
\label{N2pt_1}
\avg{ N^\pm_\ph(P) N^\pm_\ph(P') } &= \frac{ie^2}{4\pi}  \ln \frac{\L}{\mu} \coth \s . 
\end{split}
\end{equation}
This is precisely the two-point function of a massless scalar field in $H_3$ with Neumann boundary conditions ($\D=0$). In addition, we also have
\begin{equation}
\begin{split}
\avg{ N^+_\ph(P) N^-_\ph(P') } &=0 . 
\end{split}
\end{equation}
This follows from the fact that the full Maxwell action has the form $S|_{\CA_+} + S|_{\CA_-} + S |_\CD$ and so the $N^+$ and $N^-$ modes decouple from each other. Altogether, we have
\begin{equation}
\begin{split}
\label{N2pt}
\avg{ N^\eta_\ph(P) N^{\eta'}_\ph(P') } &= \frac{ie^2}{4\pi} \d_{\eta,\eta'} \ln \frac{\L}{\mu} \coth \s . 
\end{split}
\end{equation}

To end this section, we present an alternative (but related) derivation of \eqref{Na_der} and \eqref{N2pt}. We start with \eqref{action_Anu} and separate the integral over $\nu \in {\mathbb R}$ into a piece $\nu\in[-\e,\e]$ and a separate piece $\nu \in (-\infty,-\e) \cup (\e,\infty)$. We are interested in the contribution of the zero mode, which is
\begin{equation}
\begin{split}
S^\pm &\sim \frac{1}{e^2}  \int_{-\e}^\e \frac{\dt \nu}{2\pi} \! \int_{H_3} \!\! \dt^3 P \! \left(  - \frac{1}{4\nu^2} |{\tilde F}^\pm_{ab}|^2 + \frac{1}{2}  |{\tilde A}^\pm_a|^2 \right)  .
\end{split}
\end{equation}
As we take $\e \to 0$, the action is singular due to the double pole at $\nu=0$. This can be fixed by requiring the $\nu=0$ mode to be pure gauge. This implies \eqref{Na_der}. Thus, for this mode the mass term becomes the new kinetic term,
\begin{equation}
\begin{split}
S^\pm &\sim \frac{\e}{ 2\pi e^2}  \int_{H_3} \dt^3 P ( \p_a N^\pm_\ph )^2 .
\end{split}
\end{equation}
This is the action of a massless scalar on $H_3$ and its connected two-point function (with Neumann boundary conditions) is
\begin{equation}
\begin{split}
\avg{ N^\pm_\ph(P) N^\pm_\ph(P') } = \frac{ie^2}{4\e} \coth \s . 
\end{split}
\end{equation}
This is precisely \eqref{N2pt_1} if we identify $\frac{1}{\e} = \frac{1}{\pi}  \ln \frac{\L}{\mu}$.

\section{Dressing the Coulomb Phase}
\label{sec:Nphdef}

In this section, we show that the zero mode introduced in the previous section also contributes to the Wilson line dressing \eqref{Wilson_Line}. As always, we use the decomposition \eqref{A_decomp} and this time, we focus on the contribution of ${\hat A}$ to \eqref{Wilson_Line}. This is
\begin{equation}
\begin{split}
i Q_i \int_{\g_i} {\hat A} = i Q_i \int_{-\infty}^\infty \dt \tau\, {\dot y}^a(\tau) {\hat A}_a(\tau , P(\tau) ) .
\end{split}
\end{equation}
$P^\mu(\tau)$ parameterises the worldline $\g_i$ in hyperbolic coordinates. Note that if we take $\g_i$ to be the straight-line path $y^a(\tau) = y^a_i = $ constant, then the Wilson line above vanishes. This is expected, since the Coulomb phase dressing captures precisely the deviation of the particle worldline due to Coulombic interactions with the other particles.

To evaluate this, we therefore proceed as follows. We consider the path $\g_i$ that starts at the origin $x^\mu=0$, follows a straight-line trajectory to a reference point $(\pm\infty,P_\star)$ on $i^\pm$, and then moves along $i^\pm$ from $(\pm\infty,P_\star)$ to $(\pm\infty,P_i)$. The straight-line part of $\g_i$ does not contribute to the Wilson line. The path at infinity contributes and we find
\begin{equation}
\begin{split}
\label{Wilson_Line_phase}
i Q_i \int_{\g_i} {\hat A} = i \eta_i Q_i \left[ N^{\eta_i}_\ph(P_i) - N^{\eta_i}_\ph(P_\star)\right] .
\end{split}
\end{equation}
The appearance of the reference point $P_\star$ is necessitated by the additive shift ambiguity \eqref{N_identification}. In section \ref{sec:irdiv}, we showed that the Wilson dressing associated with the Goldstone exactly reproduced the real infrared divergences. It turns out, however, that the same does not hold for the infrared divergent phase. This is due to the appearance of the reference point in \eqref{Wilson_Line_phase}. To fix this, we propose a modified dressing given by
\begin{equation}
\begin{split}
i {\tilde \Phi}^{\eta_i}_i = i \eta_i Q_i \left[ N^{\eta_i}_\ph(P_i) - \avg{ N^{\eta_i}_\ph(P_\star) } \right]  . 
\end{split}
\end{equation}
All we have done is replace the operator $N^{\eta_i}_\ph(P_\star)$ with its one-point function $\avg{ N^{\eta_i}_\ph(P_\star)  }$. Note that our dressing is invariant under the additive shift \eqref{N_identification}, and is therefore well-defined.

Now that we have constructed the dressing, we can use the two-point function \eqref{N2pt} and \eqref{Gaussian_Property} to show that
\begin{equation}
\begin{split}
\label{ph_phase_corr}
\avg{ \prod_i {:} e^{i {\tilde \Phi}^{\eta_i}_i} {:} } = e^{ - i \Im \G_\ph } 
\end{split}
\end{equation}
which is the second equation in \eqref{U_decomp_prop}. Note that we have included a normal-ordering to define the exponential operator. This is required here as the two-point function \eqref{N2pt} is singular as $\s \to 0$. The same normal-ordering is not required for the real dressing since that involves the exponential of a smeared operator, $C^\pm(P)$.

\section{Infrared-Finite Fock Amplitudes in Gravity}
\label{sec:gravity}

The construction of the zero mode in perturbative gravity exactly mirrors the construction in gauge theory. We very briefly review this here. 

The infrared-divergent soft factor \eqref{soft_factorisation_ID} for gravity is given by 
\begin{equation}
\begin{split}
\label{Gamma_gr_def}
\G_\gr &= \frac{i \ka^2}{8} \sum_{ij} \eta_i \eta_j \int_\mu^\L \frac{\dt^4\ell}{(2\pi)^4} \\
&\qquad \quad \times \frac{ ( p_i \cdot p_j )^2 - \frac{1}{2} m_i^2 m_j^2 }{  ( \ell^2 - i\e ) (  p_i \cdot \ell - i \eta_i \e ) (  - p_j \cdot \ell - i \eta_j \e )} , 
\end{split}
\end{equation}
where $\ka = \sqrt{32 \pi G_N}$. This evaluates to
\begin{equation}
\begin{split}
\label{Gamma_gr_calc}
\G_\gr &= - \frac{\ka^2}{64\pi^2} \ln \frac{\L}{\mu} \sum_{ij}  \eta_i \eta_j m_i m_j \frac{\s_{ij} \cosh ( 2 \s_{ij} ) }{ \sinh \s_{ij} } \\
&\qquad \quad - \frac{i \ka^2}{64\pi}  \ln \frac{\L}{\mu}  \sum_{i \neq j} \d_{\eta_i,\eta_j} m_i m_j \frac{\cosh ( 2 \s_{ij} ) }{ \sinh \s_{ij} }  .
\end{split}
\end{equation}

$\Re \G_\gr$ is reproduced by an operator $\tilde R_{\gr,i}^{\eta_i}$ that is qualitatively similar to the one in QED (see \cite{Ware:2013zja, Choi:2017bna} for details). Gravitational theories are invariant under diffeomorphisms which act as
\begin{equation}
\begin{split}
\label{h_gauge_transform}
h_{\mu\nu}(x) \to h_{\mu\nu}(x) + \p_\mu \xi_\nu(x) + \p_\nu \xi_\mu(x) . 
\end{split}
\end{equation}
As in gauge theories, it is convenient to decompose this into a ``radiative part'' ${\hat h}_{\mu\nu}(x)$ and a pure gauge mode $C_\mu(x)$, 
\begin{equation}
\begin{split}
\label{h_decomp}
h_{\mu\nu}(x) = {\hat h}_{\mu\nu}(x) + \p_\mu C_\nu(x) + \p_\nu C_\mu(x) , 
\end{split}
\end{equation}
where ${\hat h}_{\mu\nu}$ is defined in terms of the linearised Riemann tensor as 
\begin{equation}
\begin{split}
\label{gauge_cond_h}
{\hat h}_{\mu\nu}(x) =  - 2 x^\rho x^\s \int_0^1 \dt s\, s ( 1 - s ) R^{(1)}_{\mu\rho\nu\s}(sx) .
\end{split}
\end{equation}
The pure gauge mode here is defined up to global Poincar\'e transformations,
\begin{equation}
\begin{split}
C_\mu(x) \sim C_\mu(x) + a_\mu + \o_{\mu\nu} x^\nu . 
\end{split}
\end{equation}
The gauge transformation \eqref{h_gauge_transform} then acts as
\begin{equation}
\begin{split}
{\hat h}_{\mu\nu}(x) \to {\hat h}_{\mu\nu}(x) , \qquad C_\mu(x) \to C_\mu(x) + \xi_\mu(x) .
\end{split}
\end{equation}
Small diffeomorphisms ($\xi_\mu \to 0$ at infinity) are redundancies in our description of the system, but large diffeomorphisms are physical symmetries known as BMS supertranslations and superrotations \cite{Bondi:1962px, Sachs:1962wk, Barnich:2011ct}. The Ward identities associated with these symmetries are the leading and subleading soft-graviton theorems, respectively \cite{Strominger:2013jfa, He:2014laa, Cachazo:2014fwa}. Supertranslations are spontaneously broken by the Fock vacuum down to the four global translations and the associated Goldstone mode is denoted by $C_\gr({\hat q})$. The effective action for this was constructed in \cite{Kalyanapuram:2021bvf, Kapec:2021eug} and its two-point function is 
\begin{equation}
\begin{split}
\label{C_2pt_gr}
\avg{ C_\gr({\hat q}) C_\gr({\hat q}') } = \frac{\ka^2}{16\pi^2}\ln \frac{\L}{\mu} ( - {\hat q} \cdot {\hat q}' ) \ln ( - {\hat q} \cdot {\hat q}' ) .
\end{split}
\end{equation}
This is the two-point function for a two-dimensional scalar field with $\D = -1$. In terms of this, the real dressing is given by \cite{Kapec:2021eug, Choi:2026dhz}
\begin{equation}
\begin{split}
\label{real_dressing_1_gr}
 \tilde R^{\eta_i}_{\gr,i}  = - i \eta_i m_i \int \dt^2 {\hat q}\, \CK_3 ( P_i , {\hat q} ) C_\gr({\hat q}) .
\end{split}
\end{equation}
It is then straightforward to verify using \eqref{Gaussian_Property} and \eqref{C_2pt_gr} that the correlator of $e^{\tilde R^{\eta_i}_{\gr,i}}$ reproduces the real part of $\G_\gr$,
\begin{equation}
\begin{split}
\avg{ \prod_i e^{ {\tilde R}^{\eta_i}_{\gr,i}}} = e^{-\Re \G_\gr} . 
\end{split}
\end{equation}

We next turn to the dressings that reproduce $\Im \G_\gr$. As in gauge theory, these are constructed out of zero modes $N^\pm_\gr(P)$. To describe these, we start with the Pauli-Fierz action for a massless spin-2 field,
\begin{equation}
\begin{split}
\label{action_grav}
S &=  \frac{1}{\ka^2} \int_{M_4}  \dt^4 x \left(  - \frac{1}{2} ( \p_\rho h_{\mu\nu} )^2  +  ( \p^\mu h_{\mu\nu} )^2 \right. \\
&\left. \qquad \qquad \qquad \qquad \quad + \frac{1}{2} ( \p_\mu h )^2  - \p_\mu h \p_\nu h^{\mu\nu} \right) .
\end{split}
\end{equation}
As with gauge theories, we move to the regions $\CA_\pm$ and use the decomposition \eqref{h_decomp} to simplify. In the coordinates $(\tau,y^a)$, \eqref{gauge_cond_h} implies ${\hat h}^\pm_{\tau\tau} = {\hat h}^\pm_{\tau a} = 0$ and
\begin{equation}
\begin{split}
\label{hab_origin}
{\hat h}^\pm_{ab}(\tau) &= \CO(e^{4\tau}) \qquad \qquad \text{as $\tau \to - \infty$,} \\
{\hat h}^\pm_{ab}(\tau) &= e^\tau N_{ab}^\pm + \CO(1) \quad \text{as $\tau \to + \infty$.}
\end{split}
\end{equation}
The leading large $\tau$ behaviour follows from the scale symmetry \eqref{action_grav} under which $x^\mu \to \l x^\mu$ and $h_{\mu\nu} \to \l h_{\mu\nu}$. In the coordinates $(\tau,y^a)$, scale transformations take the form $\tau \to \tau + a$. Scale symmetry then implies that ${\hat h}_{ab}^\pm = \CO(e^\tau)$.\footnote{More generally, a spin-$s$ field in $D$ spacetime dimensions transforms as $\phi_{\mu_1 \cdots \mu_s} \to \l^{s+1-D/2} \phi_{\mu_1 \cdots \mu_s}$. Thus, a scalar field in four-dimensions behaves as $\phi = \CO(e^{-\tau})$ and a vector field behaves as $A_\mu = \CO(1)$. This is consistent with the results of \cite{Cheung:2016iub} and with \eqref{asymp_cond}.}

The zero mode $N_{ab}^\pm$ can also be extracted directly from the Riemann tensor as
\begin{equation}
\begin{split}
N_{ab}^\pm = 2 \int_{\mathbb R} \dt \tau\, e^{-\tau} R^{\pm(1)}_{\tau a \tau b} . 
\end{split}
\end{equation}

We now move to Fourier space
\begin{equation}
\begin{split}
\label{hbat_Fourier}
{\hat h}^\pm_{ab} (\tau ) =   \int_{\mathbb R} \frac{\dt \nu}{2\pi i} \frac{ e^{ ( 1 + i \nu ) \tau}  }{\nu - i 0^+} {\tilde h}^\pm_{ab} (\nu) .
\end{split}
\end{equation}
The boundary condition \eqref{hab_origin} implies that the Fourier transform of ${\hat h}^\pm_{ab} (\tau )$ has a simple pole at $\nu=i0^+$ and we have incorporated this in \eqref{hbat_Fourier}. We have ${\tilde h}^\pm_{ab} (0) = N^\pm_{ab}$. Plugging \eqref{hbat_Fourier} into the action \eqref{action_grav} and simplifying, we find
\begin{equation}
\begin{split}
\label{action_grav_ads}
S^\pm  &= - \frac{1}{\ka^2} \int_{\mathbb R} \frac{\dt \nu}{2\pi} \frac{1}{\nu^2} \int_{H_3} \dt^3 P \left( \frac{1}{2} | D_c {\tilde h}^\pm_{ab} |^2  \right. \\
&\left. - | D^b {\tilde h}^\pm_{ab} |^2 - \frac{1}{2} | D_a {\tilde h}^\pm |^2 + \Re\,( \p_a {\tilde h}^{\pm*} D_b {\tilde h}^{\pm ab} ) \right. \\
&\left. -  | {\tilde h}^\pm_{ab}|^2  - \frac{1+\nu^2}{2} \left( | {\tilde h}^\pm_{ab} |^2 - | {\tilde h}^\pm |^2 \right) \right)   ,
\end{split}
\end{equation}
where $D_a$ is the metric-compatible covariant derivative on $H_3$. \eqref{action_grav_ads} is the action of a massive spin-2 field in $H_3$ with mass $m^2 = - 1 - \nu^2$ \cite{Buchbinder:2000fy, Nenmeli:2026qaz}. Using the AdS/CFT dictionary, $\D(\D-2) = m^2$, we find that the boundary operator has spin-2 and scaling dimension $\D = 1 \pm i \nu$, where again the choice of sign depends on the choice of boundary conditions on $\p H_3$.\footnote{The massive spin-2 action depends on a free parameter $\xi$ (this is the higher-spin analogue of the term $\xi R \phi^2$ for a scalar field). Here, we have assumed the standard choice $\xi=1$. For $\xi \neq 1$, we have $m^2 = - 1 - \nu^2 + 4 ( 1 - \xi ) $, but the dictionary is also modified to $\D(\Delta-2)+4(1-\xi) = m^2$, so we still have $\D = 1 \pm i \nu$.}

The effective action for the zero mode ${\tilde h}^\pm_{ab}(0) = N_{ab}^\pm$ can be obtained just as we did for Maxwell theory. We note that the kinetic term in the action (and part of the mass term) diverges near $\nu=0$. This can be fixed by taking $N_{ab}^\pm$ to be of a specific form so that the kinetic term vanishes at $\nu=0$. As in gauge theories, this specific form is typically associated with a gauge symmetry of the action. Now, it is well known that when $m^2=0$ ($\nu^2=-1$), \eqref{action_grav_ads} has a gauge symmetry associated with diffeomorphisms on $H_3$, $\d_\xi h_{ab}^\pm = D_a \xi_b + D_b \xi_a$. The case we are interested in, however, is $m^2=-1$ ($\nu=0$). It turns out that for this special value of the mass, the action \eqref{action_grav_ads} admits another gauge symmetry, $\d_f h_{ab}^\pm = ( D_a D_b - g_{ab} ) f$ \cite{DESER1984396}. Using this, it is clear that the divergence in \eqref{action_grav_ads} at $\nu=0$ can be removed by setting\footnote{The gauge symmetry at $\nu^2=-1$ ($\D=0,2$) is related to superrotations whereas the gauge symmetry at $\nu^2=0$ is related to supertranslations ($\D=1$).}
\begin{equation}
\begin{split}
\label{Ngr_DDgN}
N_{ab}^\pm = - 2 ( D_a D_b - g_{ab} ) N^\pm_\gr . 
\end{split}
\end{equation}
Plugging this form into \eqref{action_grav_ads} and simplifying, we find that the effective action for $N^\pm_\gr$ is
\begin{equation}
\begin{split}
\label{Ngr_action}
S &\sim - \frac{4\e}{\pi\ka^2}  \int_{H_3} \dt^3 P \left[ ( D_a N^\pm_\gr )^2 + 3 ( N^\pm_\gr )^2  \right] .
\end{split}
\end{equation}
This is the action of a scalar field with $m^2 = +3$. We can then determine the two-point function with Neumann boundary conditions ($\D=-1$) to be
\begin{equation}
\begin{split}
\label{N2pt_gr}
\avg{ N^\eta_\gr(P) N^{\eta'}_\gr(P') } &= - \frac{i\ka^2}{32\pi} \d_{\eta,\eta'} \ln \frac{\L}{\mu} \frac{\cosh (2\s)}{\sinh \s} , 
\end{split}
\end{equation}
where, as before, we have regulated $\frac{1}{\e} = \frac{1}{\pi} \ln \frac{\L}{\mu}$. We have also included a factor of $\d_{\eta,\eta'}$ since the $N^+$ and $N^-$ modes decouple.\footnote{Equivalently, one may first compute the two-point function of $\tilde h_{ab}^\pm$ from the action \eqref{action_grav_ads}, and then obtain the two-point function of $N_{ab}^\pm$ from its coefficient of the double pole at $\nu=0$. This shows that the modes $N_{ab}^\pm$ can be expressed in terms of a scalar field as in \eqref{Ngr_DDgN}. Extracting the scalar two-point function then gives precisely \eqref{N2pt_gr}, with the regulator $\d_\text{reg}(0)=\frac{1}{2\pi}\ln\frac\Lambda\mu$.}

The contribution of this zero mode to the gravitational dressing (up to terms that depend on the reference point $P_\star$ which drop out of the final result) is 
\begin{equation}
\begin{split}
\label{dressing_phase_gr}
{\tilde \Phi}_{\gr,i}^\pm = - \eta_i m_i N^\pm_\gr (P_i) .
\end{split}
\end{equation}
Then, using \eqref{Gaussian_Property} and \eqref{N2pt_gr}, we find that correlators of $e^{ i {\tilde \Phi}_{\gr,i}^{\eta_i} }$ reproduce the imaginary part of $\G_\gr$, 
\begin{equation}
\begin{split}
\avg{ \prod_i {:} e^{ i {\tilde \Phi}^{\eta_i}_{\gr,i}} {:} } = e^{-i \Im\G_\gr} . 
\end{split}
\end{equation}
Once again, we have used a normal-ordered exponential here since the two-point function \eqref{N2pt_gr} is singular as $\s \to 0$.

\section{Dressing for Massless Particles}
\label{sec:massless}

The dressings introduced in the previous sections are defined only when the particles are massive. In this section, we explain how we can modify the dressings when one or more particles involved in the scattering amplitude are massless.

Consider the case where $m_i \to 0$. In this limit $p_i^\mu \to q_i^\mu = \o_i {\hat q}_i^\mu$, where we parametrise the null momentum $q_i$ in terms of the energy $\o_i$ and the direction ${\hat q}_i^\mu$. From its definition, we see that the rescaled vector \eqref{Pi_def} diverges as $P_i^\mu \to \frac{\o_i}{m_i} {\hat q}_i^\mu$. In the embedding formalism, this is the standard limit which maps a bulk point $P_i$ in AdS to a boundary point ${\hat q}_i$ with the conformal factor being $z_i = m_i/\o_i$. The massless dressing can then be obtained by taking the usual extrapolate limit of our bulk AdS operators.

Let us do this for gauge theories first. The dressing is given by
\begin{equation}
\begin{split}
U_i &= \exp \left( i \eta_i Q_i \int  \dt^2 {\hat q}\, \CK_2 ( P_i , {\hat q} ) C_\ph ({\hat q}) \right)   \\
& \qquad \qquad \qquad \qquad \qquad {:} \exp \left( i \eta_i Q_i N_\ph^{\eta_i}(P_i)  \right) {:} .
\end{split}
\end{equation}
For the first term in the exponential, we note that the boundary limit of the bulk-boundary propagator \eqref{Bb_def} is given by
\begin{equation}
\begin{split}
\label{KD_limit}
\CK_\D \left( P_i , {\hat q} \right) &~ \to ~ z_i^{2-\D} [ \d^{(2)} ( {\hat q}_i , {\hat q}) + \cdots ]  \\
&\qquad \qquad + \frac{\D-1}{\pi} \left[ \frac{z_i^\D}{( - 2 {\hat q}_i \cdot {\hat q} )^\D } + \cdots \right] ,
\end{split}
\end{equation}
where the $\cdots$ denote subleading terms in $z_i=m_i/\o_i$. For the second term in the exponential, we recall that $N^\pm_\ph(P)$ is a bulk scalar operator with $\D=0$. The associated boundary operator is then given by the extrapolate dictionary as
\begin{equation}
\begin{split}
\label{Nph_limit}
N^\eta_\ph({\hat q}) = \lim_{z \to 0} N^\eta_\ph(P). 
\end{split}
\end{equation}
The boundary two-point function can also be obtained by applying the extrapolate dictionary,
\begin{equation}
\begin{split}
\label{Nph_2pt_bdy}
\avg{ N^\eta_\ph({\hat q}) N^{\eta'}_\ph({\hat q}') } &= \frac{ie^2}{4\pi} \d_{\eta,\eta'} \ln \frac{\L}{\mu} . 
\end{split}
\end{equation}

Using \eqref{KD_limit} and \eqref{Nph_limit}, we then find that in the massless limit the dressing reduces to
\begin{equation}
\begin{split}
\label{Ui_gauge_massless}
U_i &=  e^{ i \eta_i Q_i C_\ph ({\hat q}_i)} \, {:} e^{  i \eta_i Q_i N_\ph^{\eta_i}({\hat q}_i) } {:} \, .
\end{split}
\end{equation}
In particular, we see that in the massless limit, the operator $U_i$ reduces to a simple boundary operator. Using \eqref{C_2pt} and \eqref{Nph_2pt_bdy}, it is easy to show that this dressing correctly reproduces the infrared divergence in gauge theories.

It should be noted that in a gauge theory with charged massless particles, the structure of infrared divergences is more complicated than the one presented in \eqref{Gamma_ph_calc}. This is due to collinear divergences that arise in addition to soft divergences, both of which contribute to the infrared-factorisation of the amplitude. Part of this can be fixed by replacing the first exponential operator in \eqref{Ui_gauge_massless} by a normal-ordered exponential that removes the singularities that occur when ${\hat q}_i \to {\hat q}_j$. This, however, does not by itself give a complete treatment of the collinear sector. We leave a more systematic discussion of this issue for future work.

In gravitational theories, the dressing is
\begin{equation}
\begin{split}
U^\gr_i &= \exp \left(   - i \eta_i m_i \int \dt^2 {\hat q}\, \CK_3 ( P_i , {\hat q} ) C_\gr({\hat q})  \right) \\
&\qquad \qquad \qquad \qquad {:} \exp \left( - i \eta_i m_i N^{\eta_i}_\gr (P_i) \right) {:} .
\end{split}
\end{equation}
The massless limit of the first term in the exponential can be determined using \eqref{KD_limit}. For the second term in the exponential, we recall that $N^\pm_\gr (P)$ is a bulk scalar operator with $\D=-1$. The boundary operator is then defined as
\begin{equation}
\begin{split}
\label{Nph_limit_gr}
N^\eta_\gr({\hat q}) = \lim_{z \to 0} z N^\eta_\gr(P). 
\end{split}
\end{equation}
The boundary two-point function is
\begin{equation}
\begin{split}
\label{Ngr_2pt_bdy}
\avg{ N^\eta_\gr({\hat q}) N^{\eta'}_\gr({\hat q}') } &= - \frac{i\ka^2}{16\pi} \d_{\eta,\eta'} \ln \frac{\L}{\mu} ( - {\hat q} \cdot {\hat q}' ) . 
\end{split}
\end{equation}
Using \eqref{KD_limit} and \eqref{Nph_limit_gr}, we find that the massless limit of the gravitational dressing is
\begin{equation}
\begin{split}
U^\gr_i =  e^{ - i \eta_i \o_i C_\gr({\hat q}_i) } \,  {:} e^{ - i \eta_i \o_i N^{\eta_i}_\gr ({\hat q}_i) } {:} \, .
\end{split}
\end{equation}
Using \eqref{C_2pt_gr} and \eqref{Ngr_2pt_bdy}, it can be verified that this dressing correctly reproduces infrared divergences in gravitational theories.

Note that the two-point functions \eqref{Nph_2pt_bdy} and \eqref{Ngr_2pt_bdy} of the boundary operators $N^\pm_\ph$ and $N^\pm_\gr$ differ from the two-point functions \eqref{C_2pt} and \eqref{C_2pt_gr} of their counterparts $C_\ph$ and $C_\gr$ simply by the replacement $\ln ( - {\hat q} \cdot {\hat q}' ) \to - i \pi  \d_{\eta,\eta'} $. This suggests that when all particles are massless, the Coulomb dressing can be absorbed into the Goldstone dressing at the cost of modifying the Goldstone two-point function by the replacement $\ln ( - {\hat q} \cdot {\hat q}' ) \to \ln ( - {\hat q} \cdot {\hat q}' ) - i \pi \d_{\eta_i,\eta_j}$. This was observed previously in \cite{Gonzo:2022tjm}. However, this prescription only works when the amplitude involves at most one massive particle in the in- or out-state.

\section{Summary and Conclusions}
\label{sec:conclusions}

In this work, we have argued that the failure of FK dressed states to factorise into a product of one-particle states is not a  fundamental feature of gauge and gravitational theories, and have demonstrated that  when expressed in terms of the appropriate soft modes, the Hilbert space of dressed states again admits a tensor product structure like that of a Fock space. The central ingredient in this construction is a novel zero mode. By analysing the structure of the gauge and gravitational field within the light cone, we have identified a gauge-invariant zero-energy degree of freedom that is distinct from the Goldstone mode associated with large gauge transformations and supertranslations. The two-point function of this zero mode exactly reproduces the imaginary part of the infrared divergence, namely the Coulomb phase.

This fills an important gap in the literature. While it is known that the Goldstone modes reproduce the real part of the infrared divergence, the corresponding construction for the Coulomb phase has been less clear. For massless matter, the Coulomb phase can be accounted for through a shift prescription in the Goldstone two-point function \cite{Gonzo:2022tjm}. To the best of our knowledge, however, no construction in terms of soft degrees of freedom has been available for the Coulomb phase in the general massive case. Our analysis shows that the missing ingredients are the gauge-invariant zero modes $N^\pm_\ph$ and $N^\pm_\gr$. Moreover, the massless limit of our result correctly reproduces the shift prescription. This provides a derivation of the prescription from first principles.

We close by discussing a few directions for future work. First, the field $C_\ph(P)$ on $H_3$ that appears in the massive dressing is not an independent degree of freedom, but is related to the Goldstone mode $C_\ph(\hat q)$ on the celestial sphere by a bulk-to-boundary propagator, as in \eqref{CPi_def}. It would be interesting to understand whether the new zero mode $N^\pm_\ph(P)$ admits a similar description in terms of the soft-photon mode $N^\pm(\hat q)$ on the celestial sphere, which would identify it as the symplectic partner of the Goldstone mode. If this is the case, then the massive construction developed here may provide a useful framework for studying the symmetry interpretation of simultaneous double soft limits. The sum of two generic null momenta is timelike. Thus, the total momentum associated with a generic simultaneous soft limit naturally defines a point on $H_3$. The zero mode $N^\pm_\ph$ may provide the appropriate ingredient for identifying the symmetries associated with such limits.

Second, the soft modes that appear in our construction correspond to fields on $H_3$ at special values of the mass ($m^2=0$ for gauge theory and $m^2=-1$ for gravity). At these special points, we find that the action exhibits an enhanced symmetry. It would be interesting to understand this structure further, and in particular to explore whether it can be related to the infinite-dimensional symmetry algebras that appear in celestial holography, such as $w_{1+\infty}$.

It would also be useful to make the relation to the traditional FK dressings more explicit. The dressings constructed here reproduce the same infrared divergences as the FK dressings, so there should be a relation between the two descriptions. For the real part of the dressing, such a relation is already understood in terms of Goldstone modes. Our results suggest that the new zero modes are the corresponding variables for the Coulomb phase. It would be interesting to see whether an appropriate extension of the soft phase space makes the equivalence with the phase part of the FK dressing manifest.

The tensor-product factorisation of the dressed Hilbert space also allows one to revisit questions that were obscured by the non-factorisability of the traditional description. In particular, it would be very interesting to formulate an LSZ-like reduction formula directly for these dressed states. One would also like to understand how the dressed amplitudes factorise into lower-point amplitudes when an internal momentum goes on shell.

Finally, it would be instructive to extend this construction to higher dimensions. Although the logarithmic infrared divergences of four dimensions are absent in higher-dimensional gauge and gravitational theories, the underlying soft structure remains. It is therefore natural to ask whether the zero modes identified here have analogues in higher dimensions.

\section*{Acknowledgements}

We would like to thank James Drummond, Nava Gaddam, Temple He and Andrea Puhm for useful discussions. SC is supported by the European Research Council (ERC) under the European Union's Horizon 2020 research and innovation programme (grant agreement No 852386). SC is supported by the European Research Council under the European Union's Seventh Framework Programme (FP7/2007-2013), ERC Grant agreement ADG 834878. This work was supported by the Simons Collaboration on Celestial Holography. The authors used ChatGPT only for language editing of text written by authors. No scientific ideas, calculations, results, or conclusions were generated by ChatGPT. The authors reviewed all edits and take full responsibility for the manuscript.

\bibliography{FK_bib}
\bibliographystyle{apsrev4-1}

\end{document}